\documentclass [12pt,english]{article}
\usepackage{amssymb}
\usepackage{amsmath}
\usepackage{amsfonts}
\usepackage[english]{babel}

\usepackage[T2A]{fontenc}
\usepackage[cp1251]{inputenc}
\usepackage{fullpage}
\usepackage{bm}
\usepackage{graphicx}
\numberwithin{equation}{section}
\begin{document}

\title{\LARGE 
%\Roman{0}
1D half--filled paramagnetic Hubbard model. \\The Luttinger critical exponents}

\author{N. I. Chashchin\thanks{E--mail: nik.iv.chaschin@mail.ru} \\ \textit{Ural State Forestry University}\\
\textit {Ekaterinburg, 
Sibirskii trakt 37, 620100 Russia}}
\date{}

\maketitle

\begin{abstract}
The electronic system of the 1D Hubbard model is not stable due to Peierls instability; the correlations are strong even for the weak Coulomb interaction. The resulting strongly correlated state without Landau 
quisi--particle excitations is known as the Luttinger liquid. Critical exponents of a power--law 
dependence of correlation functions at low energies differ substantially for the Luttinger and Fermi liquids.  
In this paper we evaluate two critical exponents that define non--trivial behavior of the density of electronic states at low frequences and the momentum distribution of the occupation number nearby $k_F$. 
%The numerical simulation is formalized in terms of integral equations, which we obtained for the half--filled 1D %paramagnetic Hubbard model using the method of the generating functional of Green’s 
%functions and the subsequent Legendre transformation. 
\end{abstract}


%DOI:10.1134/S0031918X1607036\\
%Pacs numbers 31.15.xt; 31.15.xp; 71.10.Fd

\section{\!\!\!\!\!\!.Introduction}

Originally, the Hubbard model (HM) was introduced %\cite{Hubbard1963} 
for describing electron's correlations in narrow energy bands of transition metals. In particular, a first theoretical picture of the Mott metal--insulator transition due to one--site Coulomb electron--electron interaction had been presented. Subsequently it turned out to be very useful and convenient model for studying strongly correlated systems, their transport properties, band collapsing and mass enhancement at the Fermi level (heavy fermions) and so on. 

In last decades numerous materials appeared with a quasi one--dimensional structure such as carbon nanotubes (graphene) and organic one--dimensional conductors, where an electron's  motion over the lattice is limited by their self--interaction, i.e., correlations. The physics of interacting electrons in 1D is significantly different from that in higher dimensions. The correspondence between interacting electrons and correlated fermionic quisiparticles, which is the core of Fermi liquid (FL) theory, breaks down in 1D. The resulting strongly correlated state without Landau quisi--particle excitations known as the Luttinger liquid (LL) cannot be treated by conventional FL methods. 
Charge and spin excitations of the bosonic type become a determinal factor and specify the behavior of the system at low energies.

In 1968  Lieb and Wu (LW) published the rigorous solution of the problem based on the the Bethe ansatz 
\cite{Lieb_Wu}. 
They showed that half--filled HM has an antiferromagnetic ground state without Mott transition for any values of the 
one--site Coulomb electron–electron interaction. 
However, in this approach not all important physical parameters can be calculated,
e.g., structure factors, magnetic and charge susceptibilities, and other correlation functions. 
In particular, an important question that cannot be answered by LW solution: if HM is the Fermi liquid or the Luttinger liquid? 

All solutions of the 1D HM can be roughly divided in two main groups. The first group represents rigorous results, based on the Bethe ansatz approach, but they give almost no informations of correlation functions. 
The second group  represents solutions that were found by numerical simulations,
renormalization group approach, bosonization method. All of them allow to get some useful information
about correlation functions, but those approaches are not quite rigorous. 
 
Most impressive results were achieved by the bosonization method \cite{{Luttinger}, {Haldane1981},{Voit},{vonDelft}, {Uhrig}}. 
The gist of the method consist in rewriting an initial Hamiltonian with a linearized fermion dispersion into  pair of  uncoupled oscillators of collective charge and spin modes, and then the subsequent procedure of diagonalizing.
It was showed that the system has power-law--dependence of the correlation functions with certain critical exponents. In the spinless limit only  the one dimensionless constant K serves as an effective strength's measure  of the interaction between correlated electrons: K=1 corresponds to a non--interacting Fermi gas, K<1 to repulsion, and K>1 to attraction. The Luttinger parameter K and critical exponents are closely connected. In the spinless Tomonaga--Luttinger (TL) model \cite{Luttinger}  
the momentum distribution function and the density of one--particle states (DOS)  have the following asymptotic forms: $n(k)\propto(k_F-k)^{2\alpha}$ 
($k\approx k_F$),\, $\rho(\omega)\propto |\omega|^{2\alpha}$ ($\omega\approx 0$) 
with the same critical exponents 
\begin{equation}
\alpha = (K + K^{-1}-2)/4\,
\label{Pm:Alpa}
\end{equation} 
for both of them. 
In real 1D structures the constant K is the key parameter characterizing the behaviour of any system under considiration, and it is a quite difficult problem to get the parameter analitically \cite{Mastropietro}, numerically, and experimentally 
\cite{{Kuhne}, {Muller}}. That is because constant K is hidden in power--law--dependencies of observable theory's functions.

The purpose of our work is to get the LL solution of 1D paramagnetic HM without the bosonization scheme and the linearization of free electronic spectra; to calculate the critical exponents for the momentum distribution $n(k)$ and DOS $\rho(\omega)$, and then to estimate the Luttinger constant K for different generic Coulomb interactions. 
    
We have at our disposal an alternative vigorous instrument, namely the set of integral equations derived in our previous works  \cite{{Chaschin2011a}, {Chaschin2011b}} by the method of the generating functional of Green’s functions with the subsequent Legendre transformation.  This technique was successfully applied to the single--impurity Anderson model and the HM \cite{{Chaschin2012}, {Chaschin2016}}. As was shown, the results well agreed with the solutions obtained by other methods, and that verify the scheme.

\section{\!\!\!\!\!\!. Model and Method} 
\quad 
In the simplest form -- half--filled and symmetrical -- the Hamiltonian of the Hubbard model is written as 
\begin{equation}
\mathcal{H}= -t\sum\limits_{\langle i,j\rangle\sigma}c_{i\sigma}^{\dag}c_{j\sigma}%-\mu\sum\limits_{i\sigma}
%n_{i\sigma}
+U\sum\limits_{i}n_{i\uparrow}n_{j\downarrow}\,,
\label{Pm:Hub_ham}
\end{equation} 
where U is the parameter 
of the Coulomb interaction at a site; $c_{i\sigma}$ ($c^\dag_{i\sigma}$) annihilates (creates) an electron
with spin up and down $\sigma=\uparrow,\downarrow$; 
 $n_{i\sigma}$ is the one--site density operator; $t$ is the parameter of hopping
of electrons from site to site; in the designation
$\langle i,j\rangle$ sites are adjacent.

In our earlier articles \cite{{Chaschin2011a}, {Chaschin2011b}} we 
developed the detailed procedure of the deriving of equations for partition functions. A closed system of equations in variational derivatives determine the partition function as a functional of the inverse of free Green's functions: $Z=Z[G^{-1}_{0\uparrow},G^{-1}_{0\downarrow}]$.
Each multiparticle GF can be obtained as an multiple variational derivative of $Z$ with respect to the corresponding 
$G^{-1}_{0\sigma}$, therefore the functional $Z$ or better $\Phi=\ln Z$  is viewed as the generating functional of GFs. 

The simple iteration solution with respect to U gives well--known Feynman diagrams; the identical procedure with respect to the hopping parameter $t$ gives the series, where the atomic GF is the leading element. The follow--on Legendre transformation of those equations \cite{{Chaschin2012}, {Chaschin2016}} brings us to coupled nonlinear integral equations for propagators $N=G_\uparrow+G_\downarrow$ and  $M=G_\uparrow-G_\downarrow$. 
Here we have adapted them to the problem under consideration: the paramagnetic half--filled HM with one sub--lattice and the free electronic energy spectrum as $\varepsilon_k=- 2 t\cos(k)$ (henceforth we will suppose 2t =1).   
%\begin{equation}
%\displaystyle \varepsilon_k=-2 t\cos(k)\,.
%\label{PM_freespectr}
%\end{equation}
The momentum $k$ lies within the limits $k\in[-\pi,\pi]$, $k_F=\pi/2$, and 
the space symmetry of the model allows to make an obvious suggestion $f(k)=f(-k)$ for each function of the theory. In this case we have 
\begin{equation}
\displaystyle \frac{1}{N_{\text{at}}}\sum\limits_{k}f(k)\, = 
\displaystyle \frac{1}{2\pi}\int_{-\pi}^\pi f(k)\,d k =
\frac{1}{\pi}\int_{0}^\pi f(k)\,d k\:.
\label{PM:N_el}
\end{equation}
 
Thus, we arrive at two coupled sets of equations which are controlling correlated fermions and bosons in the system. 

The system of equations below describes fermions: 

\begin{equation}
\left\{
\begin{array}{l}
\Im N(k,\omega)=\displaystyle\frac{2\,\Im\Sigma(k,\omega)}
{\left[\omega-\varepsilon_k-\Re\Sigma(k ,\omega)\right]^2+
\left[\Im \Sigma(k;\omega)\right]^2}\,,
\\
{}
\\
\displaystyle \Im \Sigma(k,\omega) = -\frac{U}{2\pi}\int_0^\pi
\displaystyle \left[1-\tanh(\frac{\varepsilon_q}{2 T})
\displaystyle \tanh(\frac{\varepsilon_q-\omega}{2 T})\right]
\Im Q\left(q-k\;,\varepsilon_q-\omega\right) d q\,,
\\
{}
\\
\displaystyle \Re \Sigma(k,\omega) = \frac{1}{\pi}\int_{-\infty}^\infty  
\displaystyle \frac{\Im \Sigma(k,\omega^\prime)}{\omega^\prime - \omega}\,
d\omega^\prime\,;
\\
{}
\\
\displaystyle \Im\Sigma(k+\pi,-\omega) = \Im\Sigma(k,\omega), 
\quad \Re\Sigma(k+\pi,-\omega) = -\Re\Sigma(k,\omega)\,,
\\{}\\
\displaystyle \Im N(k+\pi,-\omega) = \Im N(k,\omega)\,.
\end{array}
\right.
\label{PM:Sigma_N}
\end{equation}

In Eqs. (\ref{PM:Sigma_N}) the imaginary $\Im\Sigma$ and real $\Re\Sigma$ parts of the electronic self--energy, related by means of the Kramers--Kr\"{o}nig relation, directly determine the analitical expression of the particle's number propagator $\Im N$. The last relations in the set show the useful symmetrical properties of the functions with respects to the $k$ and $\omega$. 

The bosonic charge excitations are determined by the following equiations: 

\begin{equation}
\left\{
\begin{array}{l}
\Im Q(q,\Omega)=\displaystyle\frac
{-\frac{U}{2} \,\Im \Pi(q,\Omega)}{\left[1+\frac{U}{2}\,\Re\Pi(q,\Omega)\right]^2+
\left[\frac{U}{2} \,\Im \Pi(q,\Omega)\tanh(\frac{\Omega}{2 T})\right]^2}\,, 
\\
{}
\\
\displaystyle \Im \Pi(q,\Omega) = \frac{1}{4\pi}\int_0^\pi
\displaystyle \left[1-\tanh(\frac{\varepsilon_k}{2 T})
\displaystyle \tanh(\frac{\varepsilon_k-\omega}{2 T})\right]
\Im N\left(k-q\;,\varepsilon_k-\Omega\right) d k\,,
\\
{}
\\
\displaystyle \Re \Pi(q,\Omega) = \frac{1}{\pi}\int_{-\infty}^\infty  
\displaystyle \frac{\tanh(\frac{\Omega^\prime}{2 T})\Im \Pi(q,\Omega^\prime)}{\Omega^\prime - \Omega}\,
d\Omega^\prime\,;
\\
{}
\\
\displaystyle \Im\Pi(q+\pi,-\Omega) = \Im\Pi(q,\Omega), 
\quad \Re\Pi(q+\pi,-\Omega) = \Re\Pi(q,\Omega)\,,
\\{}\\
\displaystyle \Im Q(q+\pi,-\Omega) = \Im Q(q,\Omega)\,.
\end{array}
\right.
\label{PM:Pi_Q}
\end{equation}

In Eqs. (\ref{PM:Pi_Q}) singular functions $\Pi(q,\Omega)$ and $Q(q,\Omega)$ are the charge bosonic self--energy and excitation's propagator correspondingly.
Note that both coupled sets of Eqs. (\ref{PM:Sigma_N}, \ref{PM:Pi_Q}) have the same structure; they are mathematically identical within the obvious substitutions: ${\Im N}\leftrightarrow{\Im Q}$, ${\Im\Sigma}\leftrightarrow{\Im\Pi}$, and ${\Re\Sigma}\leftrightarrow{\Re\Pi}$.   

%our master equations suitable for the numerical simulation. 
%\section{\!\!\!\!\!\!. Paramagnetic solution}

%\section{\!\!\!\!\!\!. An introduction to Luttinger liquid}

%Ordinary, multi--dimensional metals are described by Fermi liquid theory. Fermi liquid theory is a key point of 
%electron--electron interactions in metals. It states that there is one--to--one correspondence between the 
%low--energy excitations of a free Fermi gas, and those of an interacting electron liquid termed as %"quisi--particles". 
\section{\!\!\!\!\!\!. Results and Discussion}
%\addtocounter{Alpha}{1}
%{\bf\large{\Alph{Alpha}.\: Density of one-electron states and band energy--spectrums}} %\setcounter{equation}{1}
%\setcounter{figure}{0}
%\setcounter{Arab}{1}
%\setcounter{Alpha}{1} 

An elaborated computer program allowes to calculate the imaginary parts of the GFs --  $\Im N(k,\omega)$,  
$Q(q,\Omega)$, and the real parts we recieve from the Kramers--Kr\"{o}nig relations. 

Taking into account (\ref{PM:N_el}), we write  the DOS at one spin direction in a standard form:
\begin{equation}
\displaystyle \rho(\omega) =-\frac{1}{2\pi N_{\text{at}}}\sum\limits_{k}\Im N(k;\omega)=
-\frac{1}{2\pi^2}\int_{0}^\pi \,\Im N(k;\omega)\,d k\,\:.
\label{PM:rho_el}
\end{equation} 
%(\ref{PM:Sigma_N}, \ref{PM:Pi_Q}).
Graphics of the $\rho$s for U = 0, 0.5, 1.0, 2.0 are shown in Fig.1; the vestiges of van Hove singularities at 
$\omega\approx 2 t (=1)$ are noticeable even for U > 0. 
A critical exponent $\gamma$ of the function  
$\rho(\omega)\approx|\omega|^{2\gamma}$ nearby $\omega=0$ can  be estimated as 
  
\begin{equation}
\displaystyle 2 \gamma = \frac{\ln\rho(\omega)}{\ln|\,\omega|}\Bigl|_{\omega\sim 0}\,.
\label{Pm:alphar}
\end{equation} 
The calculated function $\gamma=\gamma(U)$ is depicted in Fig.4. 

\begin{figure}
\begin{center}
\includegraphics[width=0.6\textwidth]{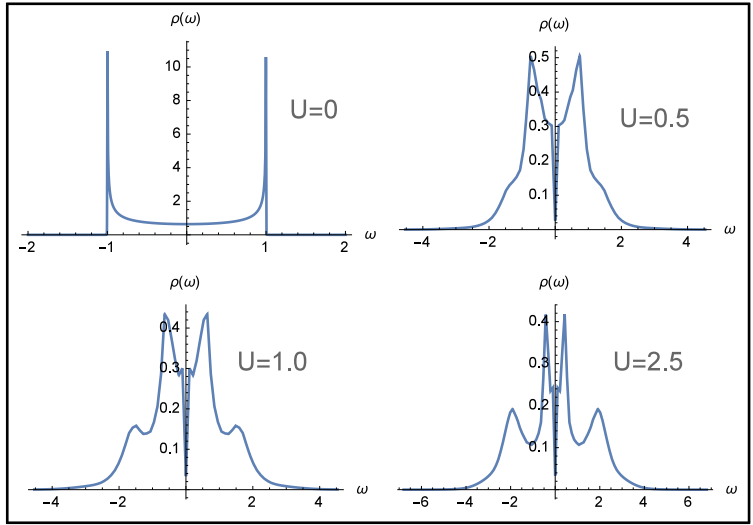}
\caption{Densities of one--electron states $\rho(\omega)$ for different U. When U increases, the energy width of states is smoothly extended, but  for U > 0 the dependence of functions at $\omega\approx 0$ is almost unchanged, and here $\rho(0)\simeq 0$.}
\label{FIG0}
\end{center}
\end{figure} 

\begin{figure}
\begin{center}
\includegraphics[width=0.6\textwidth]{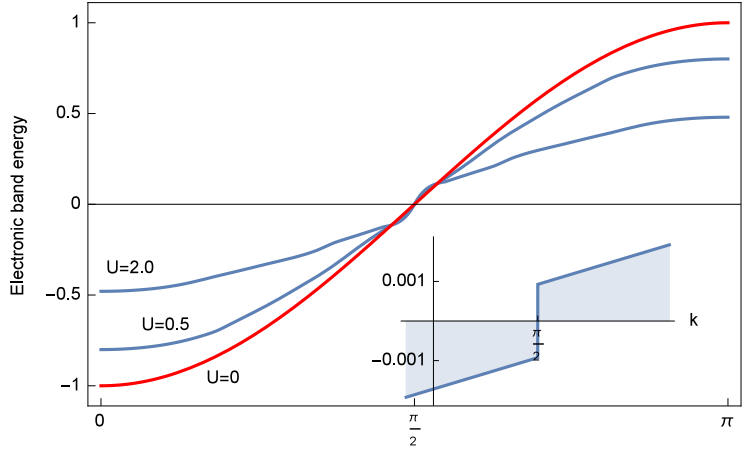}
\caption{Correlated electronic band spectra $E(k)$ for different U. As expected, on increasing U the band width  is decreasing. The inset exhibits the distinct jump at $k_F$ due to Peierls instability.}
\label{FIG1}
\end{center}
\end{figure} 

The energy spectra express a distribution of single--particle fermionic excitations in the k--space, 
$E_k=E(k)$. For strongly correlated system they display a number of visible features which distinguish  the correlated electrons  from the free ones. 
The electronic band spectrum $E(k)$ is determined from the following dispersion equation: 
\begin{equation}
\displaystyle E_k -\varepsilon_{k}-\Re\Sigma(k,E_k) = 0\,. 
\label{PM:Disp}
\end{equation}
The results of calculations we observe in Fig.2. On increasing U the band width  is decreasing, and 
the inset exhibits the distinct jump at $k_F$ due to Peierls instability, as it should be in the theory.
%\addtocounter{Alpha}{1}
%{\bf\large{\Alph{Alpha}.\: Some critical exponents of the Luttinger liquid}}

The analitical expression of the momentum distribution functions $n(k)$ is: 
\begin{equation}
\displaystyle n(k)=-\frac{1}{2\pi}\int_{-\infty}^{+\infty}f_F(\frac{\omega}{2 T})\,\Im N(k,\omega)\,d\omega\,,
\label{PM:nk}
\end{equation}
where $f_F$ is the Fermi function.  

Graphics of functions $n(k)$ for the different values U: 0, 0.5, 1.0, 2.0 are presented in Fig.3. 
The unbroken continuity of the functions \, 
$n(k)$ at $k_F$ explicitly illustrates the absence of fermionic quasi-particles; in the FL \, $n(k)$ has a jump at T=0 with  an amplitude $Z=n(k_F-0)-n(k_F+0)\ne 0$. The critical exponent $\alpha$ of the function 
$|n_{k_F}-n_k|\approx |k_F-k|^{2 \alpha}$ nearby $k=k_F$ can  be estimated as:
\begin{equation}
\displaystyle 2 \alpha = \frac{\ln[n(k)-n(k_F)]}{\ln|k-k_F|}\Bigl|_{k\sim k_F}\,.
\label{Pm:alphan}
\end{equation} 

\begin{figure}
\begin{center}
\includegraphics[width=0.6\textwidth]{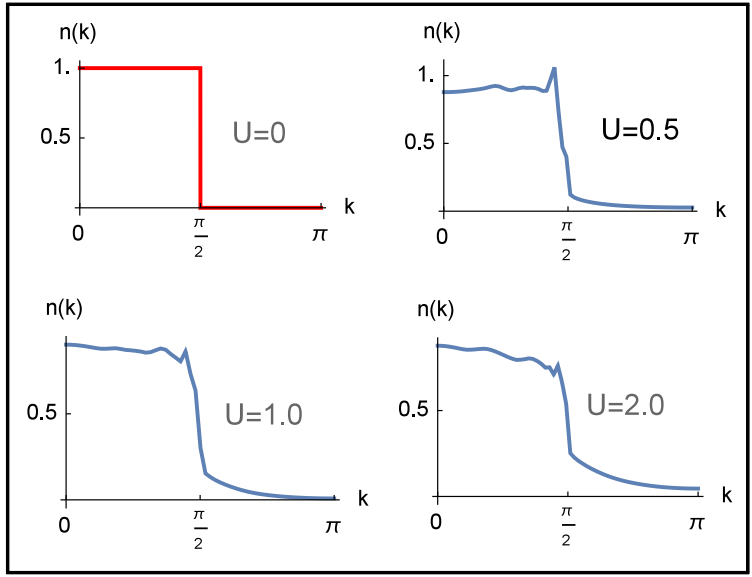}
\caption{The momentum distribution in the occupation number for different U and T=0. Due to Peierls instability there is a visible unsmoothness for $k\le k_F$. In contrast to the FL we have $Z=n(k_F-0)-n(k_F+0)=0$ for U>0. The jump is reduced to a power law functions 
$|n_{k_F}-n_k|\approx |k_F-k|^{2 \alpha}$, which implies the LL behaviour.}
\label{FIG2}
\end{center}
\end{figure}

The calculated function $\alpha=\alpha(U)$ is depicted in Fig.4. We see that the $\alpha$- and $\gamma$-critical exponents practically coincide only for very small U $\le 0.2$. Further on, their values are decreasing to $\alpha\approx$ 0.44 and $\gamma\approx$ 0.48.
Unlike the TL solution, there is a persistent difference between them, though quite small one: < 0.05. In accordance with (\ref{Pm:Alpa}) we get the Luttinger constant in the form 
 
\begin{figure}
\begin{center}
\includegraphics[width=0.6\textwidth]{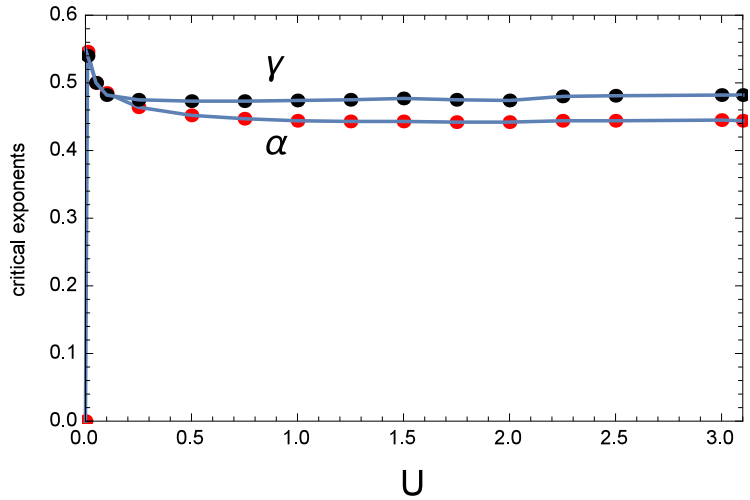}
\caption{Critical exponents versus Coulomb interaction U. Graphics $\alpha$ and $\gamma$ represent critical 
exponents for the functions $n(k\sim k_F)$ and $\rho(\omega\sim 0)$ correspondingly. Unlike the TL solution there is a difference between them for U>0.25.}
\label{FIG3}
\end{center}
\end{figure} 
 
\begin{equation}
\displaystyle K = 1+2\alpha-2\sqrt{\alpha+\alpha^2}\,.
\label{Pm:K}
\end{equation} 

Thus, we have the LL solution for all admissible Coulomb interactions, and what is more, for very small U < 0.1 the difference from the FL behaviour is more perceptible: $K\sim 0.26$ (U < 0.1), $K\sim 0.30$ (U > 1.0). Only the required paramagnetic solution $\langle (n_\uparrow-n_\downarrow)^2\rangle\ge 0$ stipulates the range of variability of Coulomb interaction $U\le 3.0$.

%\section*{Список рисунков}
%\[
%{}
%\]
%\textbf{Рис.1}\,Densities of one--electron states $\rho(\omega)$ for different U. When U increases, the energy width of states is smoothly extended, but  for U > 0 the dependence of functions near $\omega\approx 0$ is almost unchanged; everywhere $\rho(0)\simeq 0$.. 
%\[
%{}
%\]
%\textbf{Рис.2} Ряд теории возмущений для производящего функционала 
%связных \\функций Грина $\Phi=\ln Z$.
%\[
%{}
%\]
%\textbf{Рис.3} Уравнение Дайсона.

\end{document}